\documentclass[a4paper]{jpconf}
\usepackage{graphicx}
\begin{document}
\title{Shear viscosity and chemical equilibration of the QGP}

\author{Salvatore Plumari$^{1,2}$, Armando Puglisi$^{2}$, Maria Colonna$^{2}$, Francesco Scardina$^{1,2}$ and Vincenzo Greco$^{1,2}$}

\address{$^1$ Department of Physics and Astronomy, University of Catania, 
Via S. Sofia 64, I-95125 Catania (Italy)}

\address{$^2$ Laboratorio Nazionale del Sud, INFN-LNS, Via S. Sofia 63, I-95125 Catania (Italy)}

\ead{salvatore.plumari@hotmail.it}

\begin{abstract}
We have investigated, in the frame work of the transport approach, different aspects of the QGP 
created in Heavy Ion Collisions at RHIC and LHC energies. 
The shear viscosity $\eta$ has been calculated by using the Green-Kubo relation at the cascade level.
We have compared the numerical results for $\eta$ obtained from the Green-Kubo correlator with the 
analytical formula in both the Relaxation Time Approximation (RTA) and the Chapman-Enskog approximation (CE). 
From this comparison we show that in the range of temperature explored in a Heavy Ion collision the RTA 
underestimates the viscosity by about a factor of  2, while  a good agreement is found between the CE 
approximation and Gree-Kubo relation already at first order of approximation. 
The agreement with the CE approximation supplies an analytical formula that allows to develop 
kinetic transport theory at fixed shear viscosity to entropy density ratio, $\eta/s$.
We show some results for the build up of anisotropic flows $v_{2}$ in a transport approach 
at fixed shear viscosity to entropy density ratio, $\eta/s$.
We study the impact of a T-dependent $\eta/s(T)$ on the generation of the elliptic flows at both 
RHIC and LHC. We show that the transport approach provides, in a unified way, a tool able to 
naturally describe the $v_{2}(p_{T})$ in a wide range of $p_{T}$, including also the description 
of the rise and fall and saturation of the $v_{2}(p_{T})$ observed at LHC. 
Finally, we have studied the evolution of the quark-gluon composition of the plasma created in 
ultra-Relativistic Heavy Ion Collisions (uRHIC's) employing a Boltzmann-Vlasov transport approach 
that include: the mean fields dynamics, associated to the quasi-particle model, and the elastic 
and inelastic collisions for massive quarks and gluons.
Following the chemical evolution from an initial gluon dominated plasma we
predict a quark dominance close to $T_{C}$ paving the way to an hadronization via 
quark coalescence.
\end{abstract}

\section{Introduction}
The experimental results accumulated in the last decade in the ultra relativistic heavy ion collisions 
(uRHICs) before at RHIC program at BNL and more recently at the LHC program at CERN has shown that the azimuthal 
asymmetry in momentum space, namely the elliptic flow $v_{2}$, is the largest ever seen in HIC \cite{STAR_PHENIX,ALICE_2011}.
The most common approach to study the uRHICs is the viscous Hydrodynamics at second order in gradient expansion 
according to the Israel-Stewart theory \cite{Romatschke:2007mq, Molnar_cascade, Heinz}.
The comparison of the experimental measured $v_{2}$ with hydrodynamical calculations has suggested that in these 
uRHICs an almost perfect fluid with a very small shear viscosity to entropy density ratio $\eta/s$ has been 
created \cite{Romatschke:2007mq,Heinz}. Similar conclusions has been obtained also by kinetic transport theory 
\cite{Xu:2007jv,Xu:2008av,greco_cascade} .
Both Hydrodynamical and transport approach have shown an agreement on the evaluation of the viscosity with 
$4\pi \eta/s \sim 1 -3$ very close to the conjectured lower bound limit $\eta/s= 1/4\pi$. In general, both these 
calculations show that the elliptic flow depends sensitively on the ratio $\eta/s$.
Hydrodynamics however has the fundamental problem of a limited range of 
validity in $\eta/s$ and in the transverse momentum $p_T \le 2 GeV$.
At increasing $p_T$ viscous hydrodynamics breaks its validity because the relative deviation of the equilibrium 
distribution function $\delta f / f_{eq} $ increases with $p_T$ becoming large already at $p_T \geq 3T \sim 1 GeV$.
On the other hand the relativistic kinetic transport approach has the advantage to be a 3+1D approach not based on a gradient 
expansion in viscosity that is valid also for large $\eta/s$ and for out of equilibrium momentum distribution allowing 
a reliable description also of the intermediate $p_T$ range.

However usually kinetic theory is applied to the study of HIC starting from the microscopic details of the fields
and cross sections and it is not discussed directly in terms of viscosity of the system. The search for the QGP 
properties, however, have shown that the shear viscosity and in particular the viscosity to entropy density ratio 
$\eta/s$ is a key transport coefficient that could be very close to the conjectured lower bound limit, $\eta/s= 1/4\pi$.
This has lead more recently to develop a transport approach at fixed $\eta/s$ \cite{greco_cascade,Molnar_cascade,Plumari_njl}
which allows to have a direct link to the viscous hydrodynamic language.
First attempts in this direction have been already developed and applied to the study of the QGP dynamics using the 
simple expression for $\eta$ in the relaxation time approximation. On the other hand such an approach ask for a knowledge of the correct 
relation between the shear viscosity $\eta$ and temperature, cross section, mass and density.

In this proceeding we discuss different aspect of the matter created in these uRHICs within the Boltzmann-Vlasov 
transport theory.
The kinetic theory at partonic level developed solves the following relativistic equation:
\begin{eqnarray}
\label{VlasovNJL}
p^{\mu}\, \partial_{\mu} f(x,p)+M(x)\partial_{\mu} M(x) \partial_{p}^{\mu} f(x,p)=\mathcal{C}(x,p)
\end{eqnarray}
where $f(x,p)$ is the distribution function for on-shell particles and $\mathcal{C}(x,p)$ is the Boltzmann-like collision integral.
We notice that the Boltzmann-Vlasov transport theory distinguishes between the short range interaction associated to collisions 
and long range interaction associated to the field interaction, responsible for the change of the Equation of State 
(EoS) respect to that of a free gas. This last feature allows to unify two main ingredients that are relevant for 
the formation of collective flows: the Equation of State and the finite mean free path.
Furthermore the field interaction in $M(x)$ is associated to a microscopic description in terms of quasi-particles allowing to bridge
the microscopic studies of thermodynamic to the dynamics of HIC. This could potentially lead to infer more information 
on the microscopic structure of the QGP from the rich experimental observable of HIC's.
In these proceedings we will discuss a first issue related with the chemical composition of the QGP, i.e. the relative 
abundance of quarks and gluons.

The paper is organized as follows. 
In Section 2, we discuss the method for calculating the shear viscosity $\eta$ using the Green-Kubo relation
and after a brief overview of Chapmann-Enskog and Relaxation Time Approximation we show the comparison 
between our results using the Green-Kubo relation and these scheme of approximation for isotropic and 
anisotropic cross section. 
In Section 3, we use the results obtained in section 2 to introduce a transport approach at fixed shear viscosity 
to entropy density $\eta/s$ to study the build up of anisotropic flows $v_{2}$ and the effect of a temperature dependent 
$\eta/s(T)$ on the $v_{2}(p_{T})$.
In section 4, we study the evolution of the quark-gluon composition of the plasma created in uRHIC's employing a 
Boltzmann-Vlasov transport approach that include: the mean fields dynamics, associated to the quasi-particle model, 
and the elastic and inelastic collisions for massive quarks and gluons.
Finally Section 5 contains summary and conclusions.



\section{Shear viscosity from the Green-Kubo relation}
The transport coefficient like heat-conductivity, bulk and shear viscosity can be related to the correlation functions
of the corresponding flux or tensor in thermal equilibrium. The calculation of these coefficients is based on the 
fluctuation-dissipation theorem \cite{GK}. 
Here we are interested to the shear viscosity $\eta$ for which 
the Green-Kubo formula assumes the following expression \cite{zubarev}:
\begin{equation}
\label{green-kubo-completa}
\eta=\frac{1}{T} \int_{0}^{\infty}dt\int_{V}d^3x\,\langle \pi^{xy}(\textbf{x},t)\pi^{xy}(\textbf{0},t) \rangle
\end{equation}
where $T$ is the temperature, $\pi^{xy}$ is the $xy$ component of the shear component of the energy momentum tensor 
while $\langle \,...\,\rangle $ denotes the ensemble average.
In this work we determine numerically the correlation function $\langle \pi^{xy}(t)\pi^{xy}(0) \rangle$ solving the 
ultra-relativistic Boltzmann transport equation.

Our aim here is to solve numerically the full collision integral to evaluate the viscosity through the Gree-Kubo
formula and compare it with the results of the RTA and CE approximation scheme.
The particle dynamics is simulated via Monte Carlo methods based on the stochastic interpretation 
of transition \cite{Xu:2004mz,greco_cascade}.
The shear component of the energy momentum tensor is given by
\begin{equation}\label{pixy_generale}
\pi^{xy}(\textbf{x},t)=T^{xy}(\textbf{x},t)=\int \frac{d^3p}{(2\pi)^3} \frac{p^xp^y}{E}f(\textbf{x},\textbf{p};t)
\end{equation}
where we notice that at equilibrium the shear stress tensor is given by the energy-momentum tensor.
In our calculation the particles are distributed uniformly in the box. Therefore for an homogeneous system the volume averaged shear tensor
can be written as
\begin{equation}
\pi^{xy}(t)=\frac{1}{V}\sum_{i=1}^{N}\frac{p^x_i p^y_i}{E_i}
\end{equation}
the sum is over all the particles in the box.
In the Green-Kubo formula the calculation of the shear viscosity is reduced to the calculation 
of the correlation function, for details of the calculation of $\langle \pi^{xy}(t)\pi^{xy}(0) \rangle$
see \cite{Green-Kubo_2012,Wesp_2011}. Performing the numerical calculation we obtain that $\langle \pi^{xy}(t)\pi^{xy}(0) \rangle$ 
is an exponential decreasing function with the time.
We use this fact to fit the correlation function with the following expression, 
as done in several other works \cite{Wesp_2011,Fuini_3,Demir_Bass,Muronga},
\begin{eqnarray}
\langle \pi^{xy}(t)\pi^{xy}(0) \rangle = \langle \pi^{xy}(0)\pi^{xy}(0) \rangle e^{-t/\tau}
\label{corr_fit}
\end{eqnarray}
$\tau$ is the so called relaxation time.
Substituting Eq.(\ref{corr_fit}) into Eq.(\ref{green-kubo-completa}) the formula for the shear viscosity 
becomes:
\begin{equation}
\eta=\frac{V}{T}\langle \pi^{xy}(0)\pi^{xy}(0) \rangle \tau
\end{equation}
This is the formula that we will use in our calculation to extract the shear viscosity.
The relaxation time $\tau$ is calculated performing a fit on the temporal range where the correlation 
function assume the exponential form.
The error on the value of the viscosity comes from the error on the initial value 
of the correlator and the error on the relaxation time $\tau$ extracted from the fit of the correlation function.

\subsection{Green-Kubo vs Chapman-Enskog and Relaxation Time Approximation}
Before starting with the comparison between our results for $\eta$ using the Green-Kubo relation and the analytical result 
obtained in the relaxation time approximation and the Chapman-Enskog approach it is useful to show briefly these two methods.
The difference between these two methods resides in the different way in which the collision integral is approximated.
In the RTA it has been demonstrated that the shear viscosity assumes the following expression \cite{Gavin:1985ph,Kapusta_qp}:
\begin{eqnarray}
 \eta &=& \frac{1}{15T}\,\int_0^{\infty}\,\frac{d^3p_a}{(2\pi)^3}\,
\frac{|p_a|^4}{E_a^2}\,\tau_a(E_a)\,f^{eq}_a\,\,
\label{eta_relax}
\end{eqnarray}
where $\tau_a(E_a)$ is the so called relaxation time and it is related to the collision frequency of the particles.
The relaxation time can easily be expressed in terms of the total cross-section $\sigma_{tot}$:
\begin{eqnarray}
 \tau_a^{-1}(E_a)= \rho \, \langle \sigma_{tot} \,  v_{rel} \rangle
  \label{tau_eq_4}
\end{eqnarray}
We notice that in the RTA the interaction appears in the relaxation time only through the total cross-section. 
However on general physical argument the viscosity is expected to depend also on the momentum transfer that on 
average the collisions are able to produce.
In the literature to take into account this fact the relaxation time is approximated by 
$\tau^{-1}_{tr}=\langle \rho \, \sigma_{tr} \, v_{rel} \rangle$, i.e. substituting the total with the transport
cross section $\sigma_{tr}$. This is not really coming from the RTA as in Eq.s (\ref{eta_relax}) and (\ref{tau_eq_4}), however
we will refer to it as RTA in the following. As we will see such an extension of the RTA can reasonably approximate the 
correct viscosity only for the case of isotropic cross section, where it is found that $\eta_{RTA}=0.8(T/\sigma_{tr})=1.2(T/\sigma_{tot})$.

The description of the CE approach is more complicate and we use the formalism recently developed in Ref.\cite{Prakash:2012}.
At first order of approximation $[\eta_s]_{CE}^{I}$ can be written for the most general case of relativistic particles colliding with 
a generic differential cross section $\sigma(s,\Theta)$ as 
\begin{eqnarray}
 [\eta_s]_{CE}^{I} &=& 10 \,T \,\frac{{\hat{h}}^2}{c_{00}}
\label{shear_I}
\end{eqnarray}
where $\hat{h} = K_3(z)/K_2(z)$ with $z ={m}/{T}$ and $c_{00} = 16\left( w_2^{(2)} - z^{-1}\,w_1^{(2)} + (3z^2)^{-1}w_0^{(2)} \right)$.
The $w_i^{(2)}$ are the so-called relativistic omega integrals which are given by the following integral
\begin{eqnarray}
w_{i}^{(2)} &=& \frac{z^3}{K_2(z)^2}\int_{1}^{\infty} dy\, y^i 
(y-1)^7 \, K_j(2zy) \, \sigma_{tr}(y)
\label{omega_integral}
\end{eqnarray}
where $j = \frac{5}{2}+\frac{1}{2}\left( -1\right)^i$ and $y = \sqrt{s}/2M$ while 
$\sigma_{tr}=\int \, d\Omega \, \sigma(s,\Theta) \, \sin^{2}\Theta$ is the
transport cross section. 

\begin{figure}
\begin{center}
\includegraphics[width=18pc]{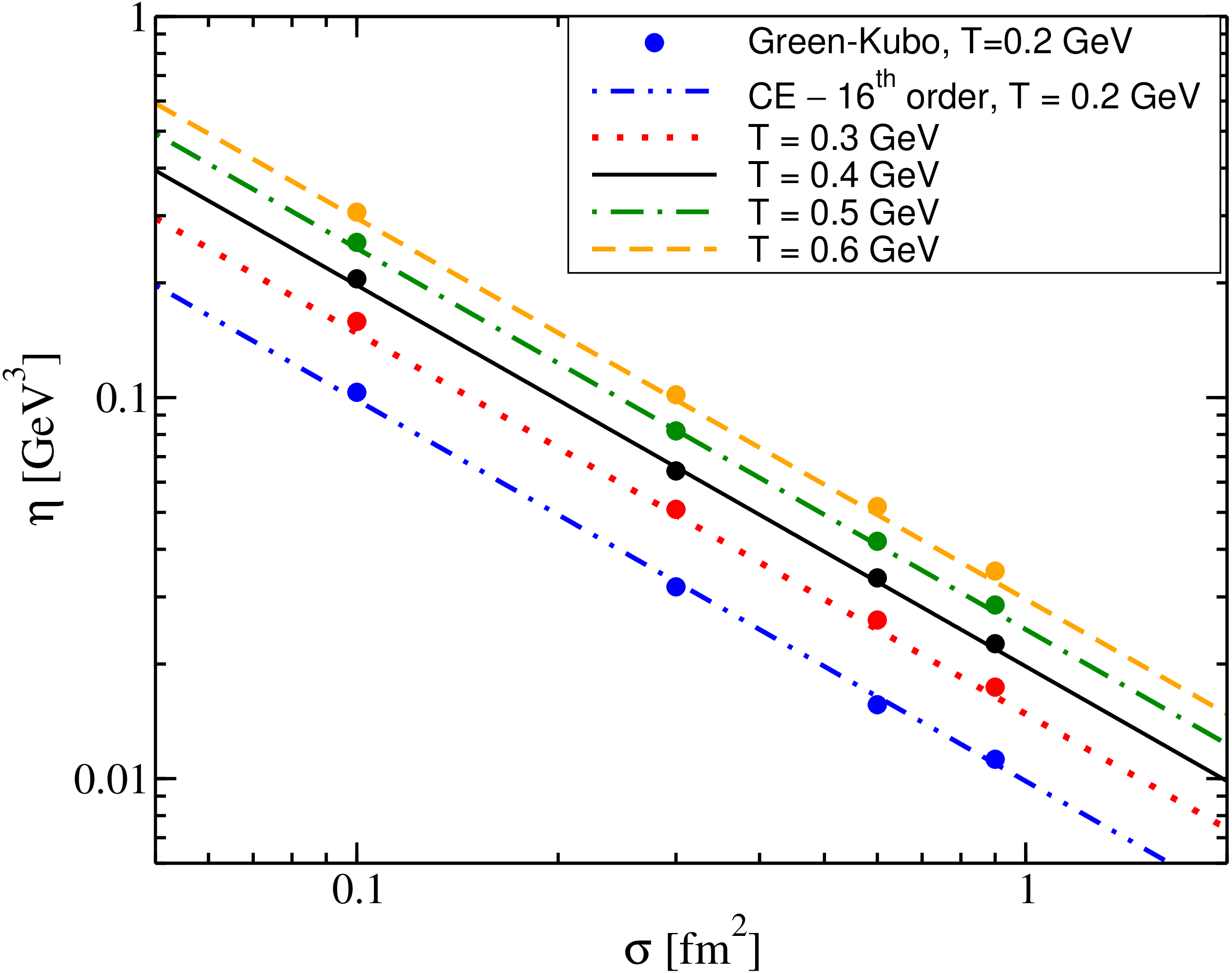}
\end{center}
\caption{\label{fig:eta_sigma}Shear viscosity $\eta$ for a massless system as a function of the total isotropic cross-section $\sigma$ and for different temperatures. The circles are the results from the Green-Kubo method, while the lines are the results obtained using the Chapman-Enskog approximation to the $16^{th}$ order Eq.(\ref{Chapman_16}).}
\end{figure}
In the following discusion we compare our results for $\eta$ using the Green-Kubo relation and the analytical result 
obtained in the relaxation time approximation and the Chapman-Enskog approach.
We perform this comparison for two different cases: first for the case of massless particles
colliding with an isotropic cross-section $\sigma(s,\Theta)=\sigma_{0}=const$ and second, for a more realistic case, 
of massless particles colliding with an anisotropic cross-section. For a more general discussion including also the 
case of massive particles see \cite{Green-Kubo_2012}.
For this simple case of isotropic cross section and massless particle the CE approximation for the shear viscosity 
is given by the following relation:
\begin{eqnarray}
[\eta]^{I}_{CE} = 0.8 \frac{T}{\sigma_{tr}}= 1.2 \frac{T}{\sigma_{tot}}
\label{Chapman_1}
\end{eqnarray}
As we can see it is the same result obtained using the RTA with $\tau^{-1}_{tr}=\langle \rho \, \sigma_{tr} \, v_{rel} \rangle$. 
In the literature there exist also higher order calculation up to the most recent work in Ref. \cite{Prakash:2012} 
where the calculation was extended up to the $16^{th}$ order. They higher order approximation converge to the value
\begin{eqnarray}
[\eta]^{16^{th}}_{CE} = 0.845 \frac{T}{\sigma_{tr}}= 1.267 \frac{T}{\sigma_{tot}}
\label{Chapman_16}
\end{eqnarray}
where however the difference between the $I^0$ and the $16^{th}$ order is about $6 \%$.
In Fig.(\ref{fig:eta_sigma}) we show the results obtained using the Green-Kubo formula
by full circles compared to the prediction of CE at $16^{th}$ order, Eq.(\ref{Chapman_16}) by lines.
The error bars for the Green-Kubo calculation are small and within the symbols.
As we can see we have a very good agreement with the analytical results in all the examined range of 
cross sections and temperatures with a discrepancy of about $2\%$. 
\begin{figure}
\begin{center}
\includegraphics[width=20pc]{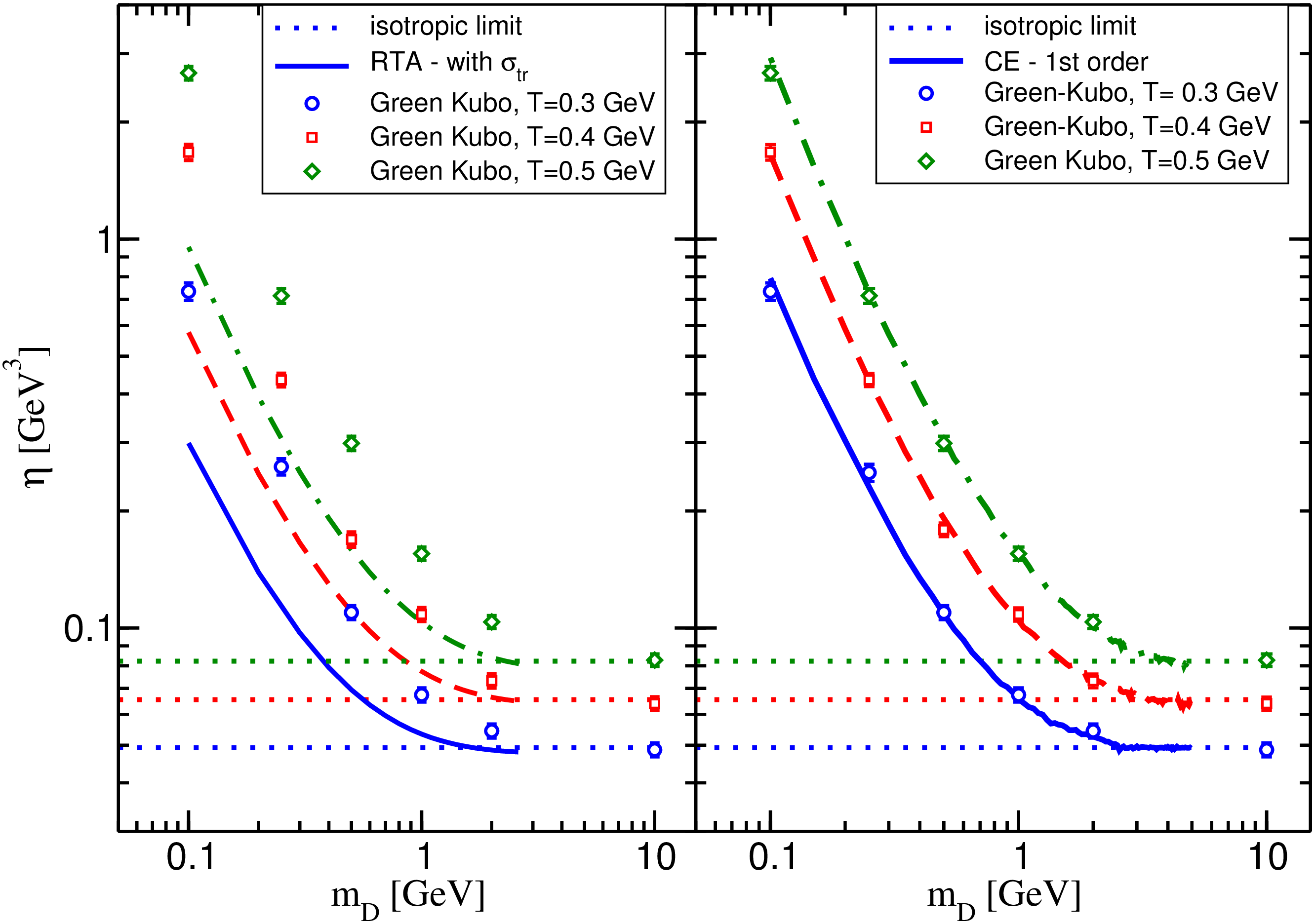}
\end{center}
\caption{\label{fig:eta_md} Left: Shear viscosity $\eta$ as a function of the Debye mass $m_{D}$ for three different values of the temperature $T=0.3, 0.4, 0.5 \, GeV$ 
blue, green and red respectively. The open simbols are the results obtained using the Green-Kubo relation.
The solid, dashed and dot dashed lines refer to the RTA approach with $\tau^{-1}=\langle \rho \, \sigma_{tr} \, v_{rel} \rangle$ respectively for $T=0.3, 0.4, 0.5 \, GeV$ . 
The dotted line is the isotropic limit when $m_D \to \infty$. 
Right: The same as the left panel but the solid, dashed and dot dashed lines refer to the Chapman-Enskog approximation at firt order.}
\end{figure}

To study the more realistic case of angular dependent cross section,we choose typical elastic pQCD inspired cross section 
with the infrared singularity regularized by  Debye thermal mass $m_D$:
\begin{equation}
 \frac{d\sigma}{dt} = \frac{9\pi \alpha_s^2}{2}\frac{1}{\left(t-m_D^2\right) ^2}\left(1+\frac{m^2_D}{s}\right) 
\label{sigma_md}
\end{equation}
where $s,t$ are the Mandelstam variables. 
Such kind of cross sections are those typically used in parton cascade approaches 
\cite{Zhang:1999rs,moln02,greco_cascade,Plumari:2010fg,Xu:2004mz,Xu:2008av}. 
The total cross-section corresponding to Eq. (\ref{sigma_md}) is $\sigma_{tot}=9\pi \alpha_s^2/(2 m_{D}^2)$ 
which is energy and temperature independent if $\alpha_{s}$ and $m_{D}$ are fixed. 
In Eq.(\ref{sigma_md}) the Debye mass $m_D$ is a parameter that regulates the anisotropy of the 
scattering cross section. 
For this case of anisotropic cross section the situation is more complicate and it is not possible to have
a simple analytical expression for the viscosity.
In the left panel of Fig.(\ref{fig:eta_md}) it is shown the shear viscosity $\eta$ as a function of the Debye mass at fixed total 
cross-section $\sigma_{tot}=3 \,mb$ and for three different temperatures $T=0.3 \, \rm GeV$, 
$T=0.4 \, \rm GeV$ and $T=0.5 \, \rm GeV$. 
On the left panel of Fig.(\ref{fig:eta_md}), we compare the Green-Kubo results (symbols) with 
the prediction of the modified RTA with $\tau^{-1}=\langle \rho \, \sigma_{tr} \, v_{rel} \rangle$ 
at first order (lines).
It is evident that there is a strong disagreement
between the two as soon as $m_D$ is far from the isotropic limit indicated by dotted lines.
Therefore, we see in general that even if the total cross section is fixed, non-isotropic cross section
strongly enhance the value of the viscosity $\eta$ and such an enhancement is stronger respect to 
$\eta \propto \sigma_{tr}^{-1}$, i.e. the modified RTA approximation. 
On the right panel of Fig.(\ref{fig:eta_md}), we compare the Green-Kubo results (symbols) with 
the prediction of CE at first order (lines). In this case we find
a very good agreement between the two, hence the CE already at first order is able to account
for the correct value of $\eta$ even if the cross section is so forward-backward peaked to cause
an increase of about an order of magnitude respect to isotropic limit.
The RTA approximation would severely underestimate the viscosity.
We can also see that for $m_D \sim 8-10 T $ the isotropic limit is recovered and both CE and RTA
coincide but this is essentially the limit discussed above.

This study and in particular the agreement obtained beetwen the Green-Kubo calculation and the CE approximation supply 
a way to develop a transport theory with fixed $\eta/s$, as we will see in the next section.

\section{The parton cascade at fixed $\eta/s$}
In this section we introduce a transport approach at fixed $\eta/s$ ratio. The motivation to 
introduce this approach is twofold: first because it is possible to make a direct comparison 
to viscous hydrodinamic approach and second, more generally, we have a tool to directly estimate 
the viscosity of the plasma valid in a wider range of $\eta/s$ and $p_{T}$ respect to hydro. 
To this end we do not calculate the cross section from a microscopic model but determine
the local cross section $\sigma_{tot}$ in order to have the wanted local viscosity. 
Here we illustrate the procedure considering the reduction of the transport approach in to a 
cascade where we neglect the field interaction in Eq.(\ref{VlasovNJL}).
In other words we consider the simplest case of massless gas.

Our approach is a $3 + 1$ dimensional Montecarlo cascade \cite{greco_cascade} for on-shell 
partons based on the stochastic interpretation of the collision rate discussed in 
Ref. \cite{Xu:2004mz}.
In the following discussion we use the pQCD inspired cross section used in the 
previous section Eq.(\ref{sigma_md}).
In the CE approximation for anisotropic cross section it is not possible to express the shear viscosity $\eta$ in an 
analytical form, because the differential cross section enters in the so called relativistic $\omega$ integrals Eq.(\ref{omega_integral}).
For this particular case $\eta$ can be written in the following form:
\begin{equation}
[\eta]_{CE}^{I}=g(m_{D},T)\frac{T}{\sigma_{tot}} \nonumber
\end{equation}
$g(m_{D},T)$ is a function of temperature and the thermal Deby mass. 
Therefore considering that the entropy density for a massless gas is $s=\rho (4 - \mu/T)$, 
$\mu$ being the fugacity, we get:
\begin{equation}
[\eta]_{CE}^{I}/s=\frac{g(m_{D},T)}{\rho (4 - \mu/T)} \frac{T}{\sigma_{tot}}\label{eq:eta_s}
\end{equation}
where $\sigma_{tot}$ is the total cross section. In our approach we solve the relativistic 
Boltzmann equation with the constraint that $\eta/s$ is fixed during the dynamics of the
collisions in a way similar to \cite{Molnar_1}
but with an exact local implementation as described in detail in \cite{greco_cascade}. 
We can evaluate locally in space and time the strength of 
the cross section $\sigma_{tot}(\rho,T)$ needed to have $\eta/s$ at the wanted value 
by mean of the following formula:
\begin{equation}
\sigma_{tot}=\frac{g(m_{D},T)}{\rho (4 - \mu/T)} \frac{T}{[\eta]_{CE}^{I}/s} \label{eq:eta_s2}
\end{equation}
An alternative but equivalent way to look at such procedure is that we implement a total cross section of the form 
$\sigma_{tot}=K(\rho, T) \sigma_{pQCD} > \sigma_{pQCD}$ where $K$ takes into account the non perturbative 
effects responsible for that value of viscosity $\eta(T)$. Note that this approach have been shown to recover 
the viscous hydrodynamics evolution of the bulk system \cite{Molnar_cascade, greco_cascade}, but 
implicitly assume that also high $p_T$ particles collide with largely nonperturbative cross section. 

\subsection{Effect of temperature dependent $\eta/s(T)$ on $v_{2}$}
In our calculation the initial condition are longitudinal boost invariant.
The initial $dN/d\eta$ have been chosen in order to reproduce the final $dN_{ch}/d\eta(b)$ at 
mid rapidity observed in the experiments at RHIC and LHC energies.
The partons are initially distributed in coordinate space according to the 
Glauber model while in the momentum space at RHIC (LHC) the partons with $p_T \leq p_0=2 GeV$ 
($4 GeV$) are distributed according to a thermalized spectrum with a maximum 
temperature in the center of the fireball of $T_{0}=2 T_C$ ($T_{0}=3 T_C$) and 
the transverse profile $T(\vec{r})=T_{0}\big( \rho(\vec{r})/\rho(0) \big)^{1/3}$
while for $p_T > p_0$ we take the spectrum of non-quenched minijets according to standard 
NLO-pQCD calculations.
\begin{figure}
\begin{center}
\includegraphics[width=15pc]{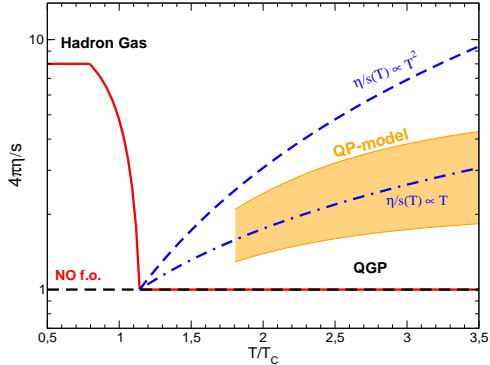}
\end{center}
  \caption{Different temperature dependent parametrizations for $\eta/s$. 
  The orange area take into account the quasi-particle model predictions for $\eta/s$ \cite{Plumari:qp_model}.}
  \label{Fig:etas_T}
\end{figure}
We also start our simulation at the time $t_0 = 0.6 fm/c$ at RHIC and $t_0=0.3 fm/c$ at LHC in agreement with the $t_{0}T \approx 1$ criterium.
In order to study the effect of the kinetic freezeout on the generation of the elliptic flow 
we have performed different calculations: one kind of calculation with a constant $4\pi\eta/s=1$ and $2$ during all the evolution 
of the system (black and green dashed line of Fig.\ref{Fig:etas_T}) the other (shown by red solid line 
in Fig.\ref{Fig:etas_T}) with $4\pi\eta/s=1$ in the QGP phase and an increasing 
$\eta/s$ in the cross over region towards the estimated value for hadronic matter $4 \pi \eta/s = 8$ 
\cite{etaS_hadronic}. Such an increase allows for a smooth realistic realization of the kinetic 
freeze-out.
\begin{figure}
\begin{center}
\includegraphics[width=20pc]{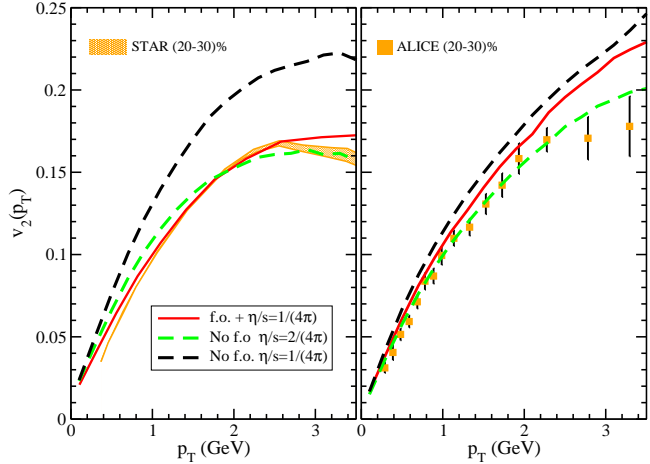}
\end{center}
  \caption{Differential elliptic flow $v_2(p_T)$ at mid rapidity for $20\%-30\%$ collision centrality. On the left panel, the orange band indicate RHIC results measured by STAR and the orange points on the right panel are the LHC results measured by the ALICE collaboration, data taken by \cite{ALICE_2011}. The black and green dashed lines are the calculations with $4\pi \eta/s = 1$ or $2$ respectively during all the evolution of the fireball and without the freeze out condition while the red line is the calculation with the inclusion of the kinetic freeze out and with $4\pi\eta/s=1$ in the QGP phase respectively.}
  \label{Fig:v2_etasT}
\end{figure}
In Fig. \ref{Fig:v2_etasT} it is shown the elliptic flow $v_2(p_T)$ at mid rapidity for 
$20\%-30\%$ centrality for both RHIC Au+Au at $\sqrt{s}=200 GeV$ (left panel) and LHC 
Pb+Pb at $\sqrt{s}=2.76 TeV$ (rhigh panel).
As we can see at RHIC energies, left panel of Fig. \ref{Fig:v2_etasT}, the $v_2$ is sensitive to 
the hadronic phase and the effect of the freeze out is to reduce the $v_{2}$ of about of $25 \%$, 
from black dashed line to red solid line and the effect of the kinetic f.o. is quantitatively similar 
to have a constant $4 \pi \eta/s = 2$ during all the evolution of the fireball, green dashed line 
in Fig. \ref{Fig:v2_etasT}. 
For the $p_{T}$ range shown we get a good agreement with the experimental data for a minimal 
viscosity $\eta/s \approx 1/(4\pi)$ once the f.o. condition is included.
This result is in agreement with the viscous hydro results, even if we notice that the agreement is good up to 
$p_{T} \sim 3 \, GeV$ and non ansantz on the non-equilibrium effect is needed.
At LHC energies, right panel of Fig. \ref{Fig:v2_etasT}, the scenario is different, we have 
that the $v_2$ is less sensitive to the increase of $\eta/s$ at low temperature in the hadronic 
phase. The effect of large $\eta/s$ in the hadronic phase is to reduce the $v_2$ by less than 
$5 \%$ in the low $p_T$ region, from black dashed line to the red solid line in right panel 
of Fig. \ref{Fig:v2_etasT}. 
This different behaviour of $v_2$ between RHIC and LHC energies can be explained looking 
at the life time of the fireball. In fact at RHIC energies the life time of 
the fireball is smaller than that at LHC energies, $4 - 5 fm/c$ at RHIC against the about $8 -10 fm/c$ 
at LHC, in agreement with HBT results. Therefore at RHIC the elliptic flow has not enough time to fully develop in the QGP phase. 
While at LHC we have that the $v_2$ develops almost completely because the fireball 
spend more time in the QGP phase.
Qualitatively these results are similar to those obtained using the formula for $\eta$ in the 
RTA (see Ref.(\cite{Plumari_Bari})) but quantitatively the results are different because as shown in the 
previous section the viscosity estimated in the RTA in the range of temperature explored in a HIC 
differes also for a factor 2 from that obtained in the CE approximation which is the correct one.
In general in these results we have a smaller $v_{2}$ respect to the results obtained used the RTA 
approximation for $\eta$ and this is due to the fact that for fixed $\eta/s$ ratio in the CE approximation 
we estimate locally a smaller total cross section respect to the RTA case.

As pointed out due to this large life time of the fireball at LHC and the larger initial temperature 
is interesting to study the effect of a temperature dependence in $\eta/s$.
In the QGP phase $\eta/s$ is expected to have a minimum of $\eta/s \approx (4\pi)^{-1}$ close to 
$T_{C}$ as suggested by lQCD calculation \cite{lQCD_eta}. While at high temperature quasi-particle 
models seems to suggest a temperature dependence of the form $\eta/s \sim T^{\alpha}$ with 
$\alpha \approx 1 - 1.5$ \cite{Plumari:qp_model}. To analyze this possible scenarios
for $\eta/s$ in the QGP phase we have considered another different with a quadratic 
dependence $4\pi \eta/s=(T/T_0)^2=(\epsilon/\epsilon_0)^{1/2}$ (green line) where 
$\epsilon_0=1.7 GeV/fm^3$ is the energy density at the beginning of the cross over regions 
where the $\eta/s$ has its minimum, see blue dashed line in Fig.\ref{Fig:etas_T}.
\begin{figure}
\begin{center}
\includegraphics[width=20pc]{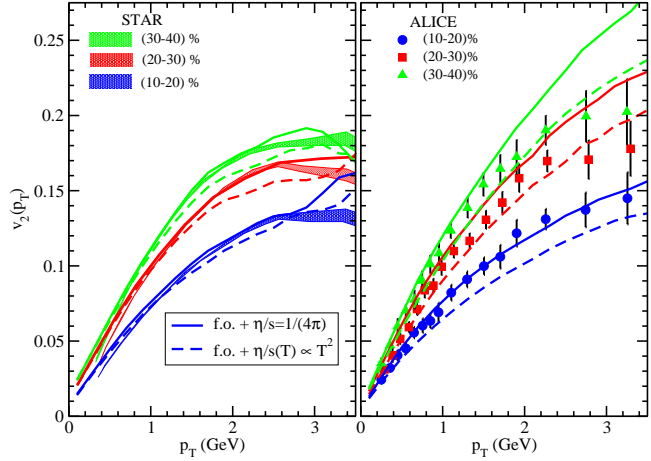}
\end{center}
  \caption{Differential elliptic flow $v_2(p_T)$ at mid rapidity for different collision centralities. On the left panel, the blue red and green bands indicate RHIC results measured by STAR and while points on the right panel are the LHC results measured by the ALICE collaboration with the same colours of the left panel for the different collision centralities, data taken by \cite{ALICE_2011}. The solid and dashed lines are calculations with the inclusion of the kinetic freeze out and with $4\pi\eta/s=1$ and $4\pi\eta/s \propto T^2$ in the QGP phase respectively. The different colour are for the different collisions centralities.}
  \label{Fig:v2_etasT2_b}
\end{figure}
At RHIC energies the $v_2$ is essentially not sensitive to the dependence of $\eta/s$ on temperature 
in the QGP phase, see the dashed blu, red and green lines in the left panel of Fig. \ref{Fig:v2_etasT2_b}.
However the effect on average is to decrease the value of $v_2$ but at low $p_T < 1.5 GeV$ 
the $v_2(p_T)$ appears to be insensitive to $\eta/s(T)$ while a quite mild dependence appears at higher 
$p_T$ where however the transport approach tends always to overpredicted the elliptic flow 
observed experimentally.
At LHC energies the build-up of $v_2$ is more affected by the $\eta/s$ in the QGP phase and 
on average it is reduced of about a $20 \%$. In any case still a strong temperature 
dependence in $\eta/s$ has a small effect on the generation of $v_2$ 
we found that with a constant or at most linearly dependent on T $\eta/s$ the transport 
approach can describe the data at both RHIC and LHC at least up to $p_T \sim 2 -3 GeV$.
The fact that we can reproduce the $v_{2}(p_{T})$ for different centralities implies that we can reproduce
also the breaking of $v_{2}/\epsilon$ scaling as described by core-corona models \cite{Becattini:2008ya,Aichelin:2008mi,Aichelin:2010ed}.
In our approach, this fact is taken into account dynamically by the increase of the $\eta/s$ ratio at lower energy density, see Fig. \ref{Fig:etas_T}.

\subsection{Effect of high $p_{T}$ partons on $v_{2}$}
In the previous section has been pointed out that in our approach where we fix locally the 
ratio $\eta/s$ we have that $\sigma_{Tot}=K(\rho,T) \sigma_{pQCD}$ therefore we have large cross 
section independently of the $p_T$ of the colliding particles.
Obviously, this procedure doesn't permit to recover the pQCD limit for hard collision 
therefore we extend our previous approach allowing for a $K$ factor that depends on the 
invariant energy of the collision $K(s)$. We choose this function in such a way that at 
high energies $K(s) \to 1$ and in this way we get the the connection between the non pertubative 
interacting bulk and the asymptotic pQCD limit. For $K(s)$ we choose the following exponential 
form  $K(s/\Lambda^2)=1 + \gamma \, e^{-s/\Lambda^2}$, where $\Lambda$ is a scale parameter that 
fix the energy scale we have a transition to pQCD behaviour.
While $\gamma$ plays the same role of $K$ in the old calculations and it is determined again 
in order to keep fixed locally the $\eta/s$. Therefore we can repeat the same procedure as 
described in the previous section but now with $\sigma_{Tot}=K(s) \sigma_{pQCD}$. 
Only for a qualitative discussion in this section we will show the results for the elliptic flow 
obtained using the relaxation time approximation (RTA) for the shear viscosity \cite{Plumari_Bari}. 
Implementation with the CE is in progress.
\begin{figure}
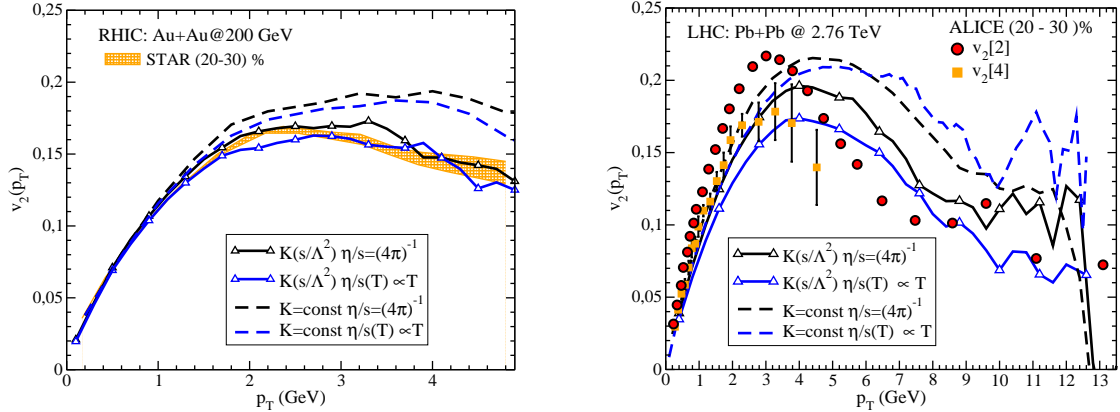

\includegraphics[width=16pc]{v2_Ks5}
  \hskip1cm
\includegraphics[width=16pc]{v2_Ks_pT5}
  \caption{Left: $v_2(p_T)$ at mid rapidity for $20\%-30\%$ collision centrality at RHIC. The dashed lines are the calculations with $K=const$ and for $4\pi\eta/s=1$ and $4\pi\eta/s \propto T$ with f.o. respectively for black and blue curves while the solid lines are the same but with $K(s/\Lambda^2)$. Right: $v_2(p_T)$ at mid rapidity and for $20\%-30\%$ collision centrality at LHC with the same legend, data taken from \cite{ALICE_2011}.}
  \label{Fig:v2_Ks_etas}
\end{figure}
In the left panel of Fig. \ref{Fig:v2_Ks_etas} we compare the $v_2(p_T)$ for $20\%-30\%$ 
collision centrality at RHIC energies with (solid) and without (dashed) the inclusion of 
an energy dependent $K$ factor and for two T dependences of the $\eta/s$. As we can see 
$K(s/\Lambda^2)$ does not affect at all the $v_2(p_T)$ for $p_T < 2 GeV$, in other words at 
RHIC energies the high $p_T$ partons do not affect the generation of the $v_2$ of the bulk. 
On the other hand, we have a reduction of the $v_2$ for $p_T > 3 GeV$ and in particular the 
$v_2$ becomes a decreasing function of $p_T$ for $p_T > 3 GeV$ in agreement with what is 
observed experimentally (orange band). In the right panel of Fig. \ref{Fig:v2_Ks_etas} we 
compare in a large range the $v_2(p_T)$ at LHC energies with (solid) and without (dashed) 
the inclusion of an energy dependent $K$ factor and for two T dependence of the $\eta/s$.
We notice that the two sets of experimental data refer to different method of $v_2$ measurements, 
namely $v_2[2]$ (circle) and $v_2[4]$ (square) and our theorethical results should be compared 
to $v_2[4]$ because event-by-event fluctuations are not considered in our calculations. As we can 
see at LHC energies the $v_2$ is sensitive to $K(s/\Lambda^2)$ already at $p_T \approx 1.5 GeV$ 
quite lower than the RHIC case, in other words the many high $p_T$ partons that we have at LHC 
energies affect the generation of the $v_2$ of the bulk. In general we observe that the $v_2(p_T)$ 
becomes more sensitive to the value of the viscosity when we include the function $K(s/\Lambda)$.

We can give the following interpretation for the $v_2(p_T)$:
The raise of the $v_2(p_T)$ at low $p_T$ is an effect of a strong interacting fluid where the particles 
interact mainly non perturbatively with large cross sections and therefore we get a description in 
agreement with hydrodynamics.
For $p_T > 3 - 4 GeV$ the elliptic flows starts to be a decreasing function of $p_T$. This is the region 
where the disappearance of the non perturbative effect significantly affects the $v_2(p_T)$ making faster 
and stronger ($\sim 20 - 25 \%$) the fall in the elliptic flow in the range $3 GeV < p_T < 8 GeV$.
Finally in our calculation for $p_T > 8 GeV$, where the pQCD limit is almost established, seems to appear 
the saturation of the $v_2$ similarly to the experimental data and typical of a path-lengh mechanism
as in jet quenching models \cite{Scardina}. In this region an analysis with better statistics is required.

\section{Chemical equilibration of the QGP}
A successful way to account for non-perturbative dynamics of the QGP is a quasi-particle approach, in which
the interaction is encoded in the quasi-particle masses 
\cite{Plumari:qp_model,Ruggieri_qpmodel,Bluhm:2010qf,LH1998,PC05,PKPS96,Ratti:2011au}.
The model is usually completed by introducing a finite bag pressure that can account for further non-perturbative 
effects. It is well known that, in order to be able to describe the main features of lattice QCD thermodynamics \cite{Borsanyi:2010cj}, 
in these quasi-particle models a temperature-dependent mass has to be considered. This also implies that the bag 
constant has to be temperature-dependent, in order to ensure thermodynamic consistency.
We notice that if the QGP can be described in terms of finite mass excitations this has a strong impact on the
quasiparticle chemical ratio $N_{q+\overline{q}}/N_g$. In fact at equilibrium one has:
\begin{eqnarray}
\frac{N_{q+\overline{q}}}{N_g}=\frac{d_{q+\overline{q}}}{d_g}
\frac{m_q^2(T)\, K_2(m_q/T)}{m_g^2(T)\, K_2(m_g/T)}\, ,
\label{eq-ratio}
\end{eqnarray}
where $K_2$ is the Bessel function and $m_{q,g}(T)$ are the $T-$dependent quark and gluon masses that can be determined by a fit 
\cite{Plumari:qp_model} to recent lQCD calculations \cite{Borsanyi:2010cj}.
In Fig.\ref{eq-ratio}, we show by solid line the equilibrium ratio when the fit to lQCD 
$\epsilon(T)$ is done assuming $m^2_q/m^2_g=3/2 \cdot (N_c^2-1)/N_c(2N_c+N_f)=4/9$
according to a pQCD scheme \cite{LH1998,PC05,PKPS96}. 

\begin{figure}[htbp]
\begin{center}
\includegraphics[scale=0.25]{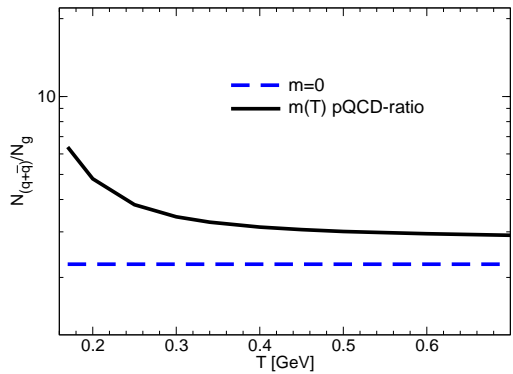}
\caption{\label{nqng-qpm}
Quark to gluon ratio at equilibrium
as a function of the temperature as predicted by QPM \cite{Plumari:qp_model}. 
For the solid line the $m_q/m_g$ ratio is according to pQCD while the dashed line indicate the massless case.}
\end{center}
\end{figure}
The motivation to use a transport equation for quasiparticles with a space-time dependent mass
resides in the success of quasi particles in describing correctly the behavior of energy density
and pressure of the QGP as computed in the lQCD approach.
We employ a Boltzmann-Vlasov transport theory to simulate the partonic stage of the HIC.
In the last years several codes have been developed based on transport theory at the cascade level
\cite{Zhang:1999bd,moln02,Xu:2004mz,greco_cascade},
i.e. including only collisions between massless partons, with quite rare exceptions
\cite{Cassing:2009vt,Plumari_njl,Scardina_2012}. 

In this section we present the results within a transport approach that includes
the mean field dynamics associated to the thermal self-energies generating the finite
mass $m(T)$ in the quasi-particle model discussed in Ref. \cite{Plumari:qp_model,LH1998,PC05,PKPS96}. 
In such a picture the relativistic Boltzmann-Vlasov equation can be written as follows:
\begin{equation}
 [p^\mu \partial_\mu+ m^*(x) \partial_\mu m^*(x)\partial ^{\mu}_p]f(x,p)=
{\cal C}[f](x,p)
\label{BV-equation}
\end{equation}
where $\mathcal{C}(x,p)$ is the Boltzmann-like collision integral, main ingredient of the several cascade
codes:
\begin{equation}
{\cal C} \!=\! 
\int\limits_2\!\!\! \int\limits_{1^\prime}\!\!\! \int\limits_{2^\prime}\!\!
 (f_{1^\prime} f_{2^\prime}  -f_1 f_2) \vert{\cal M}_{1^\prime 2^\prime \rightarrow 12} \vert^2 
 \delta^4 (p_1+p_2-p_1^\prime-p_2^\prime)
\label{coll-integr}
\end{equation}
where $\int_j= \int_j d^3p_j/(2\pi)^3\, 2E_j $, $f_j$ are the particle distribution functions,
while ${\cal M}_{f\rightarrow i}$ denotes the invariant transition matrix for elastic 
as well as inelastic processes.  
The elastic processes have been implemented and discussed in several previous
works \cite{Zhang:1999bd,moln02,Xu:2004mz,greco_cascade}.
In this section, instead, we will show some results including the inelastic processes between quarks 
and gluons ($gg \leftrightarrow q\overline{q}$) and to achieve this we have evaluated the matrix element 
in a pQCD LO order scheme. The tree diagrams contributing to the $gg \leftrightarrow q\overline{q}$ 
correspond to the $u,t,s-$channels: ${\cal{M}}={\cal{M}}_s+{\cal{M}}_t+{\cal{M}}_u$. 
For the massless case the cross sections for such processes are the textbook pQCD cross section for 
jet production in high-energy proton-proton collisions. 
In our case we have considered a finite mass for both gluons and quarks together with a dressed
gluon propagator where for vanishing gluon mass we recover the renowed Combridge cross sections 
\cite{Combridge:1978kx}. The details of the calculations are quite similar to the one in \cite{Biro:1990vj} 
for finite current strange quark mass.
The thermodynamical self-consistency of the QPM requires a self-consitency
between the Bag constant and the effective mass of the quasiparticles \cite{Plumari:qp_model}
which leads to a gap-like equation coupled to Eq.(\ref{BV-equation}):
\begin{equation}
 \frac{\partial B}{\partial m_i} =-
 \int \frac{d^{\,3} \vec{p}}{(2\pi)^3 } \frac{m_i(x)}{E_i(x)}  f_i(x,p) \, 
\label{gap-equation} 
\end{equation}
with $i=q,\overline q,g$.
Eq.(\ref{gap-equation}) allows to evaluate locally the mass in Eq.(\ref{BV-equation}) also in non-equilibrium 
conditions guaranteeing the conservation of the energy-momentum tensor of the fluid.

\begin{figure}[htbp]
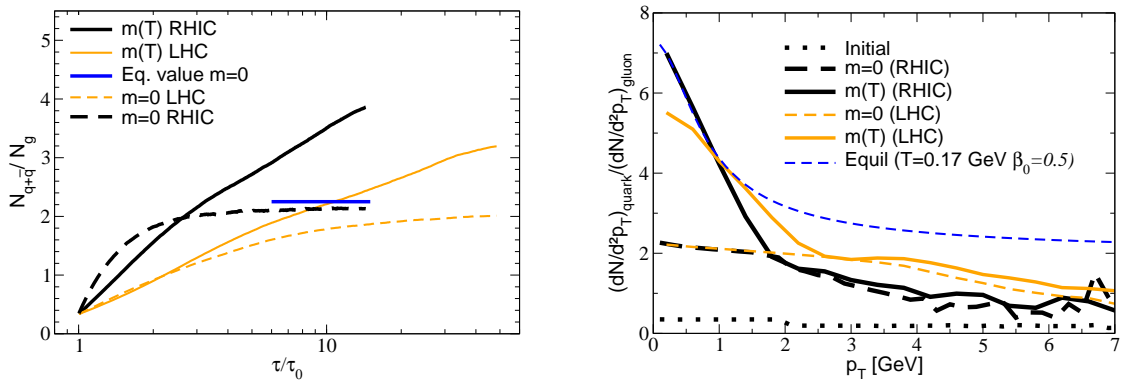

\begin{center}
\vspace{0.6cm}
\includegraphics[width=16pc]{rapp_vs_t_RHIC_LHC_xxx.eps}
  \hskip1cm
\includegraphics[width=16pc]{rapp_vs_pt_RHIC_LHC_xxx.eps}
\caption{\label{ratio-time} 
Left:Quark to gluon ratio as a function of time (normalized to $\tau_0$) in $Au+Au$ at $\sqrt{s}= 200 \rm AGeV$ 
(black lines) and for $Pb+Pb$ at $\sqrt{s}= 5.5 \rm ATeV$ (light lines). Dashed lines are for the massless case 
and the solid for the massive case.
Right: Quark to gluon ratio vs $p_T$ at the freeze-out time. The thin solid line represents the full equilibrium raio, 
Eq. (\ref{ratio-pt-anal}); see text for details.}
\end{center}
\end{figure}
We have checked that the numerical implementation in a stationary box is able to reproduce correctly 
both the kinetic and the chemical equilibrium. In particular we have carefully checked that for different 
temperature we get the correct abbundancy of quarks and gluons as given by Eq.(\ref{eq-ratio}). In the 
following discussion we will show some results for the more realistic case of a HIC.

We have simulated $Au+Au$ collisions at $\sqrt{s}= 200$ AGeV and $Pb+Pb$ at $\sqrt{s}= 5.5 \rm ATeV$
for $0-10\%$ centrality. The initial condition are longitudinal boost invariant with an initial quark 
to gluon ratio of $N_{q+\overline{q}}/N_{g}=0.25$ with about $75 \%$ of gluon and $25 \%$ of quarks.
The initial conditions in the $r$-space are given by the standard Glauber condition while in the $p$-space
we use a Boltzmann-Juttner distribution function up to a transverse momentum $p_T=2$ GeV while
at larger momenta minijet distributions are implemented, as calculated by pQCD at NLO order \cite{Greco:2003mm}. 
At RHIC the initial maximum temperature at the centre of the fireball is $T_0=340$ MeV and the initial time 
$\tau_0=0.6$ fm/c (corresponding also to the $\tau_0 \cdot T_0 \sim 1$ criterium) as in succesfull 
hydrodynamical simulations. 
For $Pb+Pb$ collisions at $\sqrt{s}= 5.5 \, \rm ATeV$ we have $T_0=600$ MeV and $\tau_0\sim 1/T_0=0.3$ fm/c.
In Fig.\ref{ratio-time} it is shown the time evolution of the ratio $R_{qg}=N_{q+\overline{q}}/N_g$
for both the massless case (dashed lines) and the massive quasi-particle case (solid lines). As we can see 
in the massless case at RHIC the system reaches very quickly, in less than 1 fm/c, the chemical equilibrium given by 
$R_{qg} \sim 2$. 
The difference in the time evolution of the ratio between RHIC and LHC is due to the temperature dependence 
of the strong coupling $\alpha_{S}(T)$ that we have considered in our model.
In fact during the first stages of the fireball evolution at LHC the temperature is larger with respect 
to the initial stages at RHIC and this means a smaller $\alpha_{S}(T)$ at LHC.
However for the final ratio we obtain similar results for RHIC and LHC and this behaviour 
is consistent with the fact that at the equilibrium this ratio does not depend on the temperature 
of the system, see dashed line in Fig.(\ref{nqng-qpm}).

When the quasi-particle massive case is considered, as we can see the ratio $R_{qg}$ is an increasing function 
with the time and it reaches $R_{qg} \sim 3.8$ for $Au+Au$.
As we can see in Fig.(\ref{ratio-time}) for the massive quasi-particle case (solid lines) we have a difference 
in the time evolution of the ratio $N_{q+\overline{q}}/N_{g}$ between RHIC and LHC and this is consistent,
not only with the temperature dependence of $\alpha_{S}(T)$, as in the massless case, but also with the fact that the 
equilibrium value is strongly $T$ dependent especially close to $T_c$ (see Fig.\ref{nqng-qpm}) 
and the system is sensitive to the different initial temperature between RHIC and LHC. 
Nonetheless we find that the fireball reaches a value relatively close to the equilibrium at $T \sim T_c$ 
and eventually it is composed by about $80 \%$ of quark plus anti-quarks.
As shown we have that the time evolution of the quark to gluon ratio $N_{q+\overline{q}}/N_{g}$ is qualitatively similar 
between RHIC and LHC but for LHC it is always smaller than that at RHIC. This difference is due to the fact that at LHC 
longer part of the lifetime is spent in a $T$-region where the equilibrium $R_{qg}$ is nearly constant. This results 
into a moderately smaller final ratio.

In the right panel of Fig.\ref{ratio-time} the $p_T$ dependence of the quark to gluon ratio is shown for 
the initial distribution (dotted line) and the freeze-out distributions: massless case (dashed line) and 
massive case (solid line). Black lines are for $Au+Au$ and light ones fore $Pb+Pb$.
We see the large difference between the massless and the massive case and also that the net gluon to quark 
conversion extends up to quite large $p_T$. 
As we can see we have that at very low $p_{T}$ this ratio is very close to the equilibrium value for both massive with 
$N_{q+\overline{q}}/N_{g} \approx 6 - 7$ and massless with $N_{q+\overline{q}}/N_{g} \approx 2$ at the f.o. where the 
temperature is $T \approx T_{C}$. The decrease of the ratio with $p_T$ can be expected considering that high$-p_T$ particles
can more easily elude the equilibration dynamics. However, in the massive case, we note a quite strong dependence 
below $p_T=2$ GeV that has not to be interpreted as a fast detachment from the chemical equilibrium.
In fact the $p_T-$dependence of the ratio can be evaluated analitically at equilibrium and it is given by
\begin{equation}
 \frac{dN/d^2\,p_T|_{q+\overline{q}}}{dN/d^2\,p_T|_g}=\frac{d_{q+\overline{q}}}{d_g}
\frac{m_{T}^{q} e^
{\gamma[(m_{T}^{q}-\beta_0 p_T)/T]}}{m_{T}^{g} e^
{\gamma[(m_{T}^{g}-\beta_0 p_T)/T]} }
\label{ratio-pt-anal}
\end{equation}
where $\beta_0$ is the radial flow velocity, $m_{T}^{q}$ and $m_{T}^{g}$
are the transverse masses.
In the right panel of Fig.(\ref{ratio-time}) with the thin dashed line we plot the function of 
Eq.(\ref{ratio-pt-anal}) rescaled by a factor 0.85 accounting for the lack of full thermalization.
The strong $p_T$ dependence obtained in the transport simulation follows very closely the equilibrium behavior
at least up to $p_T \sim 1.5$ GeV. This is a well known effect predicted by hydrodynamics and experimentally observed 
from SPS to LHC energy for hadronic spectra. 
In conclusion the effect of the mass on the chemical composition of the QGP is substatially both at high $p_{T}$ and 
low $p_{T}$ region where we have a dominance of quarks also in high $p_{T}$ region. this result represent an important 
support to the idea of a coalescence of masssive quraks \cite{Greco:2003xt,Greco:2003mm,Fries:2003kq,Fries:2008hs}.


\section{Conclusions}
In the frame work of the transport approach we have investigated different aspects of the QGP created in 
Heavy Ion collisions at RHIC and LHC energies.

We have developed a method to solve numerically the Green-Kubo formula for the shear viscosity for 
the case of a relativistic Boltzmann gas.
In our study we have compared the Chapman-Enskog approximation and the relaxation time approximation
 with the result from the Green-Kubo correlator that in principle should provide the correct result.
We have shown our results only for two physical case: isotropic and non-isotropic cross section for 
massless, for a more complete study see \cite{Green-Kubo_2012}.
Our work shows that the relaxation time approximation always underestimate the shear viscosity even 
by more than a factor of 2 and it gives a satisfying prediction only in the unrealistic case of massless 
isotropic cross section. Instead in the Chapman-Enskog approximation already at first order have shown 
an agreement at the level of $2\%$ with the numerical calculations of the Green-Kubo correlator for all 
the physical case considered.
The agreement between the CE approximation at first order and the Green-Kubo method also supplies a 
relatively simple analytical expression for the viscosity that we have used to developed a kinetic 
transport theory at fixed viscosity with very good precision. This shows that the current used relaxation 
time approximation for the viscosity can lead to significantly underestimate the $\eta/s$. Therefore 
previous works \cite{greco_cascade,Abreu:2007kv,Plumari_Bari} in this direction are only approximately valid.

We have investigated within a transport approach at fixed $\eta/s$, in the CE approximation, the effect of a 
temperature dependent $\eta/s$ at RHIC and LHC energies. An important result is that at LHC a key observables 
like the elliptic flow is much less contaminated by the hadronic phase allowing a better study of the QGP 
properties. This result is qualitatively in agreement with the previus work in the RTA.
We get for both at RHIC and LHC a good agreement with the data when the ratio $\eta/s \approx 1/(4\pi)$ 
and in general we observe not a large sensitivity of $v_2$ to the T dependence in $\eta/s$. 
Furthermore we have seen that at LHC the large ammount of particle with $p_{T} > 4 GeV/c$ 
interacting nearly perturbatively cannot be neglected.  
The interplay between perturbative and non-perturbative behaviour seems to have an important 
effect on the generation of $v_2$ at intermediate $p_T$ and it could explain the rapid raise 
and fall of $v_{2}(p_{T})$ in $0<p_{T}<8 GeV/c$ shown in the experiments.

Finally, we have studied the evolution of the quark-gluon composition of the plasma created in 
ultra-Relativistic Heavy Ion Collisions (uRHIC's) implied by standard quasi-particle approach developed to 
study the thermodynamics of the QGP. Tis has been realizd employing a Boltzmann-Vlasov transport theory 
that includes both elastic and inelastic collisions.
This study shows that one could expect that the QGP created in uRHIC's, even if it is initially a 
Glasma should very quickly evolve into a plasma dominated by quark plus antiquarks close to the 
cross-over temperature $T_c$. 
With the quark to gluon ratio can evolve by more than a factor 20 and at freeze-out is anyway almost 
a factor 2 larger than the one for an equilibrated massless QGP.
The results are quite robust and devolpments of the QPM \cite{Plumari:qp_model,LH1998,PC05,PKPS96} or inclusion of 
three-body inelastic scatterings may even make the effect larger. 
The result supplies a theorethical justification of the \textit{massive-quark} coalescence 
hadronization models able to successfully describe several puzzling observations at RHIC and LHC 
\cite{Greco:2003xt,Greco:2003mm,Fries:2003kq,Fries:2008hs}.

\section*{References}

\bibliography{iopart-num}

\end{document}